\documentclass[conference]{IEEEtran}
\IEEEoverridecommandlockouts

\usepackage[nocompress]{cite}
\usepackage{amsmath,amssymb,amsfonts}
\usepackage{algorithm}
\usepackage{algorithmic}
\usepackage{graphicx}
\usepackage[mathlines,switch]{lineno}
\usepackage{multicol,multirow}
\usepackage{marginnote}
\usepackage{stfloats}
\usepackage{color}
\usepackage{mathptmx}
\usepackage{comment}
\usepackage{ulem}
\usepackage[style=base]{caption}
\usepackage{subcaption}

\def\BibTeX{{\rm B\kern-.05em{\sc i\kern-.025em b}\kern-.08em
    T\kern-.1667em\lower.7ex\hbox{E}\kern-.125emX}}
\begin{document}

\title{AI/ML in 3GPP 5G Advanced - Services and Architecture}

\author{Pradnya Taksande\IEEEauthorrefmark{1}, Shwetha Kiran\IEEEauthorrefmark{2}, Pranav Jha\IEEEauthorrefmark{3} and Prasanna Chaporkar\IEEEauthorrefmark{4}
\\
\IEEEauthorblockA{Department of Electrical Engineering,
Indian Institute of Technology Bombay, India\\
Email: 20001816@iitb.ac.in\IEEEauthorrefmark{1},
shwethak@iitb.ac.in\IEEEauthorrefmark{2},
pranavjha@ee.iitb.ac.in\IEEEauthorrefmark{3},
chaporkar@ee.iitb.ac.in\IEEEauthorrefmark{4}
}
}

\maketitle

\begin{abstract}
The 3rd Generation Partnership Project (3GPP), the standards body for mobile networks, is in the final phase of Release 19 standardization and is beginning Release 20. Artificial Intelligence/ Machine Learning (AI/ML) has brought about a paradigm shift in technology and it is being adopted across industries and verticals. 3GPP has been integrating AI/ML into the 5G advanced system since Release 18. This paper focuses on the AI/ML related technological advancements and features introduced in Release 19 within the Service and System Aspects (SA) Technical specifications group of 3GPP. The advancements relate to two paradigms: (i)~enhancements that AI/ML brought to the 5G advanced system (AI for network), e.g. resource optimization, and 
(ii)~enhancements that were made to the 5G system to support AI/ML applications (Network for AI), e.g. image recognition.
\end{abstract}

\section{Introduction}\label{intro}
%
Artificial Intelligence (AI) and Machine Learning (ML) are transforming numerous industries and multiple aspects of modern life. From personalized recommendations on streaming platforms to real-time fraud detection in banking, AI/ML technologies are driving smarter decision-making across industries. In retail, they assist in inventory and supply chain management. In transportation, automotive vehicles rely on ML for object detection and navigation. Businesses leverage AI/ML for predictive analytics, customer insights, and operational efficiency. 
As data continues to grow, these technologies are evolving rapidly, reshaping how we work, interact, and solve complex problems, making them central to innovation in today's world. 

AI/ML is playing a transformative role in mobile networks to enhance task efficiency, optimize network management and configuration, predict fault detection, and improve end-user experience. The 3rd Generation Partnership Project (3GPP), a global body responsible for mobile network standardization, began incorporating AI/ML features starting from Release 17 of the 5th Generation (5G) system. 
The 3GPP Standards development work is divided across three major groups: Radio Access Network (RAN), Service and System Aspects (SA), and Core Network \& Terminals (CT). The SA group is responsible for standardizing Core Network (CN) functionalities, Security, Operations, Administration and Maintenance (OAM) functionality and end-to-end application support. 
The RAN group develops base station and radio interface standards, whereas the CT group develops User Equipment (UE) and CN protocols as well as interworking with external networks. Within 3GPP, AI/ML is being explored across multiple areas, including RAN intelligence, OAM intelligence, and incorporation of AI/ML capabilities in 5G Advanced architecture that began with Release 18.

In 5G and beyond, plans are underway to support AI/ML related use cases like distributed computing for edge-cloud collaboration, energy-efficient operation at network as well as UE, autonomous network management, making mobile networks more adaptive, scalable, and resilient to changing demands. These use cases can broadly be divided into:

\noindent
1)~{\bf AI for Network}: It includes the use of AI to optimize traditional algorithms, network functions, as well as network operations to improve performance, efficiency, and end-user experience. For instance, AI/ML can be used to design an efficient network resource management algorithm. 

\noindent
2) {\bf Network for AI}: Here, the network needs to provide capabilities like communication with Quality of Service (QoS) guarantees, computing and storage resources to support AI/ML applications. For instance, when running an image recognition AI/ML application on the UE, the low latency connectivity to the edge or cloud needs to be provided by the network for additional processing support.

\begin{figure*}[!ht]
\centering
\includegraphics[width=0.75\linewidth]{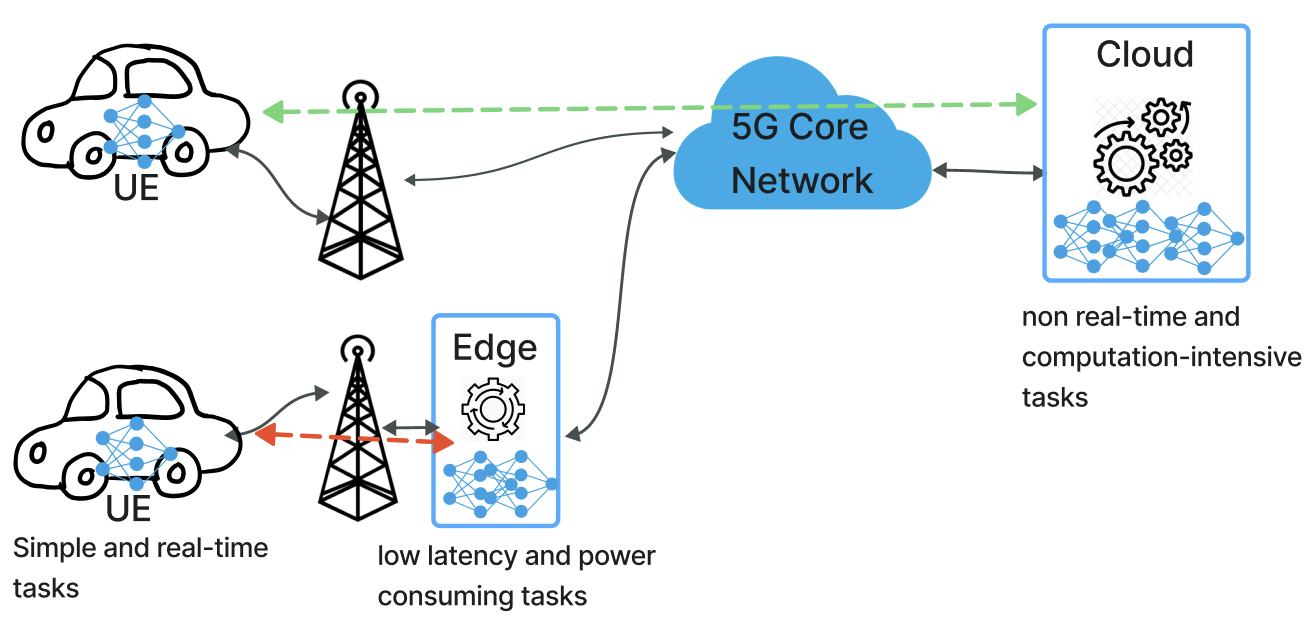}
\caption{Figure shows an autonomous car connected to 5G network. The car uses hierarchical AI/ML models. In-vehicle ML model manages real-time tasks like emergency braking and obstacle avoidance. For more complex and computation-intensive tasks, ML models are available at the edge and cloud nodes. Edge (cloud, resp.) models are used for moderate latency (non real-time, resp.) tasks like route optimization (optimization of battery management system, resp.). This hierarchical deployment ensures efficient task allocation based on latency and computational requirements. The ML model in the car may use data from its sensors, the model at the edge may use data from the cars in the same neighbourhood, and the model in the cloud can utilize global data. The data path in edge and that in cloud is shown with red and green dotted line, respectively.
} \label{car}
\end{figure*}

Number of surveys have been published summarizing potential of AI/ML for
optimizing mobile networks. A comprehensive overview of AI and its role in communications within the context of 6G networks has been provided in \cite{cui2025overview}, however, the discussion on standardization aspects of AI/ML remains at  a high level. \cite{alzailaa} and \cite{sharma}  discuss AI/ML in network contexts, particularly related to RAN, but they do not focus on standardization. \cite{yeh2024standardization} primarily focuses on RAN standardization across various standardization bodies, viz. 3GPP and O-RAN. Similarly, \cite{sun2024combination} also emphasizes the RAN perspective but does not adequately cover the AI/ML activities carried out by 3GPP SA group. \cite{qu2023overview} provides a comprehensive overview of AI/ML technologies within the 3GPP RAN scope, including the proposed framework for applying AI/ML to the New Radio (NR) air interface. \cite{lin2025bridge} provides an overview of Release 19 RAN enhancements in 5G-Advanced, purely from the 3GPP standardization perspective. 

At present, no literature specifically examines AI/ML capabilities introduced in 3GPP Release 19 from the SA group's perspective. This article aims to bridge that gap. 
In the next section, we present challenges that need to be addressed while integrating AI/ML in networks. Subsequently, we provide a detailed account of AI/ML-related enhancements introduced in Release 19 within the 3GPP SA group. Finally, we present the AI/ML related study items approved for Release 20 to provide insight into AI/ML capabilities expected to be a part of 6G. 
%

\section{AI/ML in Mobile Networks} \label{aiml_in_networks}
AI/ML is playing an increasingly important role in mobile networks enabling smarter, efficient and adaptive networks, and also in various user applications. Thus, future networks are required not only to support AI/ML for its own efficient functioning, but also to provide communication, storage and compute support to AI/ML based applications. In this regard, 3GPP standardization looks at following two aspects: 

\noindent
1. {\bf AI-centric (Network for AI):} This deals with the network efficiently supporting AI/ML functionalities like support for Federated Learning (FL), data and model exchanges between user applications and cloud servers, and providing desired QoS for communication requirements. See Fig.~\ref{car} for an illustrative example.

\noindent
2. {\bf Network-centric (AI for network):} This deals with efficient and secure acquisition, storage and management of data from various network elements, and then utilizing the data to train models that are integrated to improve networks' efficiency.


Next, we cover challenges in both scenarios mentioned above.

\noindent
1. \textit{Hierarchical computing:} Since mobile devices are usually battery-powered with limited computing and storage capabilities, processing complex tasks locally may not be feasible. To address this, the AI/ML workload can be split between the mobile device and an edge and/or cloud server.
The challenge is in deciding how to split the ML model, which nodes to use, what are the Key Performance Indicators (KPIs) for each, how much computation power is required at each node, how much energy is consumed, and so on.

\noindent
2. \textit{ML model distribution and sharing:}
The complexity of the ML model increases with the depth of the Deep Neural Network (DNN), as more layers demand more computational resources and result in increased model complexity. Moreover, the model’s performance is affected by the distribution of training data, which may change over time and differ across geographic regions, often necessitating separate models to maintain accuracy in each context. 
Selecting the right model and determining the optimal compute node is a key challenge. Due to limited storage at the UE, it may be necessary to download models from an edge/cloud on demand. 
For ML model transfer between the UE and edge/cloud, a significant amount of information must be exchanged over the network. This exchange involves not just raw data, but also computational data and ML model parameters, in both uplink and downlink directions. To handle the surge in data exchange, the network must evolve to support significantly higher demands of communication services and storage requirements. 

\noindent
3. \textit{Distributed/Federated learning:}
Distributed learning is a learning approach where the training of ML models is spread across multiple nodes. FL is a type of distributed learning where multiple nodes (like UEs or edge nodes) collaboratively train a shared model without exchanging raw data. Instead, each node trains the model locally using its own data and only shares model updates (like gradients or weights) with a central entity, which aggregates them to improve the global model. The most common aggregation approach is iterative model averaging. Each node trains the downloaded model using local data and sends interim results (e.g., gradients) to the cloud server over 5G uplink. The server aggregates these updates, refines the global model, and sends the updated model back to the nodes via 5G downlink for the next training round. 
The latency of the FL process is determined by the convergence rate (number of iterations before the training process converges to a consensus) and the computation plus communication latency of each iteration.
%
The latency and data rate during these transfers play a critical role in ensuring the accuracy of the model.

4.~\textit{Challenges in AI for network:}
Similar to the challenges of 'Network for AI' indicated above, there are also many challenges in utilizing AI/ML technology to improve network performance and operation. AI/ML-based techniques depend on large amounts of quality data to function effectively. For example, in the domain of network security, AI/ML tools need terabytes of data to build accurate models to detect threats. Similarly, huge amount of data is required to utilize ML-based resource management schemes for improved network performance. However, such data may not be available in public domain as industries are reluctant to share their network logs and sensitive data with third parties. This may be a significant problem for network operators catering to a small number of subscribers. Hence, there is a need to establish an open network dataset. To deal with the issue of data unavailability, the network may also try to use synthetic data, which again is challenging due to issues of distribution bias etc. There is also a need for AI/ML and sensing technologies to work together to improve mobile network performance, e.g, usage of network-sensed data may help in ML-based resource management schemes. However, a challenge here is that the data sensed from the radio environment may not always have the required labels to be of use. AI models can sometimes also produce false alarms or miss critical threats in network monitoring. AI/ML-based schemes also require significant processing power, resulting in increased demands on hardware and scalability.

\begin{figure*}[!ht]
\centering
\includegraphics[width=0.75\linewidth]{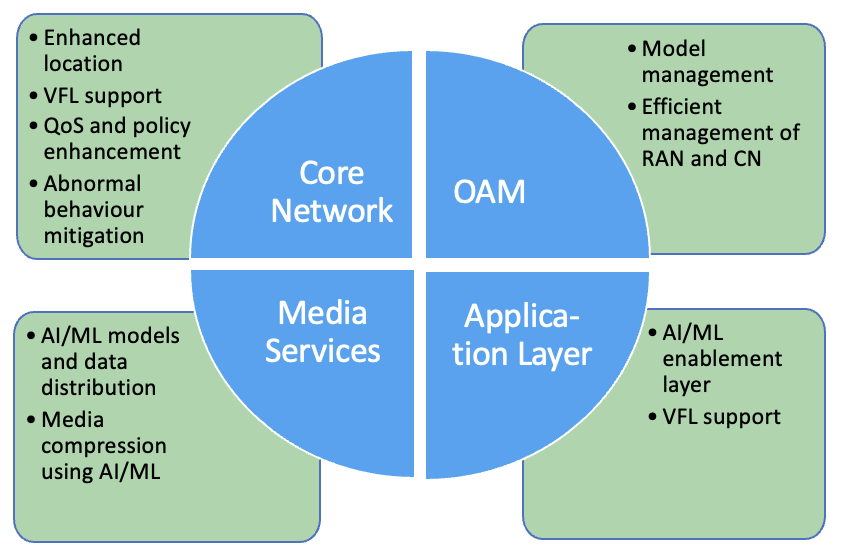}
\caption{Figure shows different elements in the 5G System and corresponding AI/ML enhancements. Blue circle shows network components that are enhanced, while summary of enhancement is given in the green box next to the network component.}
\label{features}
\end{figure*}
\begin{figure*}[!ht]
\centering
\includegraphics[width=0.75\linewidth]{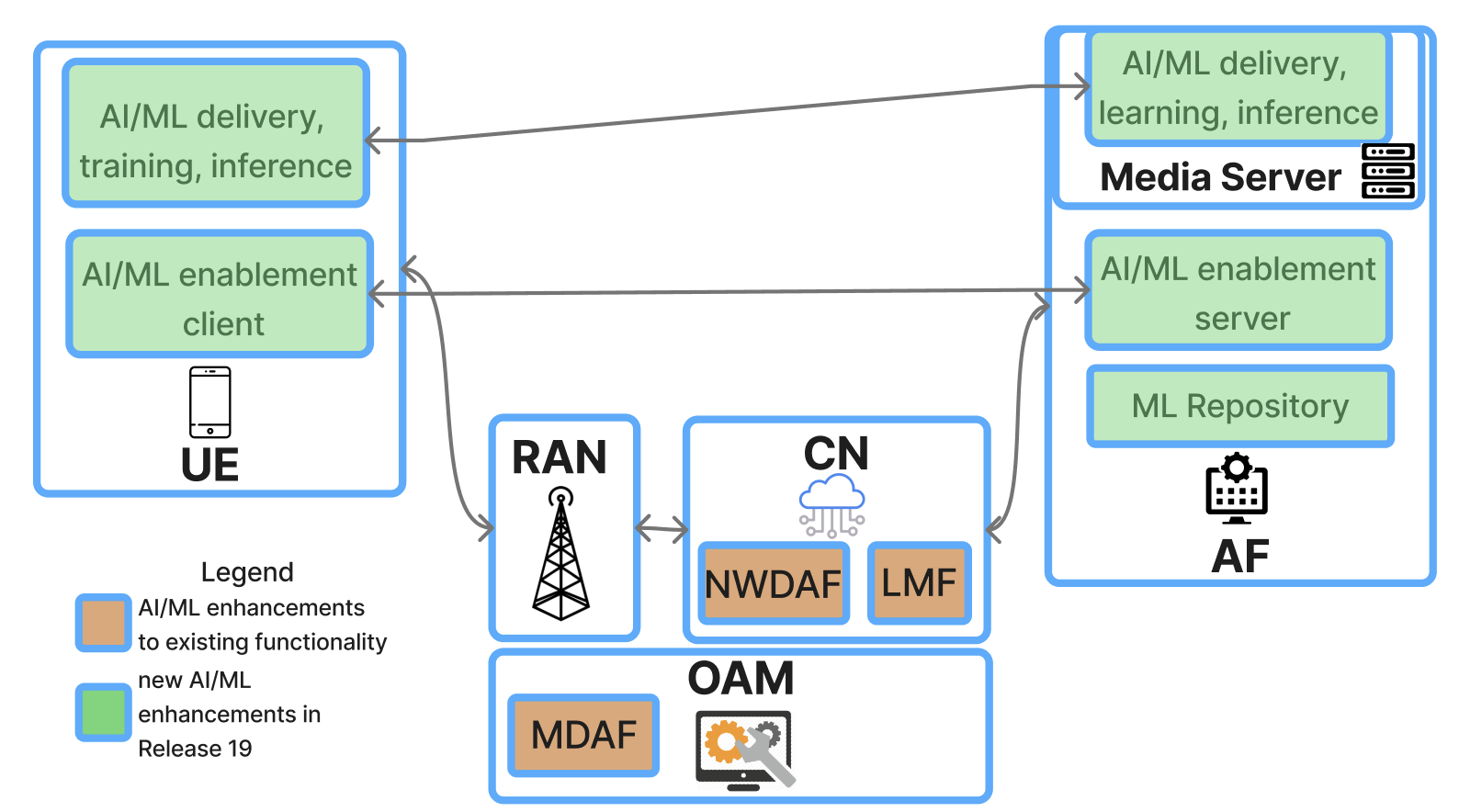}
\caption{An overview of AI/ML enhancements in 5G system in Release 19 (SA group). 
}
\label{overall}
\end{figure*}
To address the challenges mentioned above, 3GPP is defining clear guidelines and feature requirements. Features like AI-assisted coverage and capacity optimization, and beam management were introduced in RAN along with QoS and applications management in the CN. In addition, AI/ML model management to support operations and maintenance activities was also introduced. A summary of Release 19 AI/ML enhancements across different domains of the 5G system is presented in Fig.~\ref{features}.
For AI/ML related enhancements in Release~19,
Fig.~\ref{overall} provides an overview, and Table \ref{aiml_table} lists all the related study and work items. In the following sections we dive into the details of the AI/ML features defined in Release~19. 
\section{AI/ML Enhancements to 5G System} \label{sa1}
For facilitating 
ML model sharing or distribution, and FL, 3GPP introduced system requirements for model transfer and distribution using network connection in Release 18. These requirements include the experienced uplink and downlink rates for various scenarios such as split model transfer, split inference, or FL corresponding to applications like image recognition, video streaming, robotics and speech recognition \cite{ts22261}. For instance, the upper limit for split AI/ML inference and model downloading has been set at 1.1 Gbps for downlink, while in hotspot areas, it is limited to 4 Gbps. With these benchmark requirements in place, the AI/ML model can potentially prioritize and select users in good coverage areas as suitable members for FL. 

In Release 19, 3GPP introduced direct Device-to-Device (D2D) connectivity to support efficient AI/ML incorporation for a range of applications, including auto-driving, robot remote control, and video recognition. This includes support for AI/ML model transfer over 5G direct device connection for distributed learning/inference. For instance, in distributed inference scenario, a device with valuable input data but with low power/computation resource, may calculate the first few layers of an AI model and offload the intermediate data to another device using direct device connection for the final inference result. For D2D connection, 3GPP defined specific KPIs, QoS performance criteria and system requirements. For example, in split AI/ML operations between endpoints using direct device connections for AI inference, the maximum experienced data rate is capped at 1.5 Gbps, and end-to-end latency is capped at 10 msecs for factory robots applications. An instance of system requirement states that the 5G System (5GS) shall be able to dynamically add or remove specific UEs to/from the AI/ML FL task when communicating via direct device connection. 

\section{AI/ML Enhancements to 5G Core Network} \label{sa2}
%
3GPP introduced AI/ML related advancements to 5G CN as early as Release 17 with some ML functionality support in Network Data Analytics Function (NWDAF). In Release 18, the 5G system supported ML model distribution, transfer, and training for applications like voice/video recognition, robot control, automotives and so on. 
\begin{table*}[!t]
\centering
\includegraphics[width=2\columnwidth]{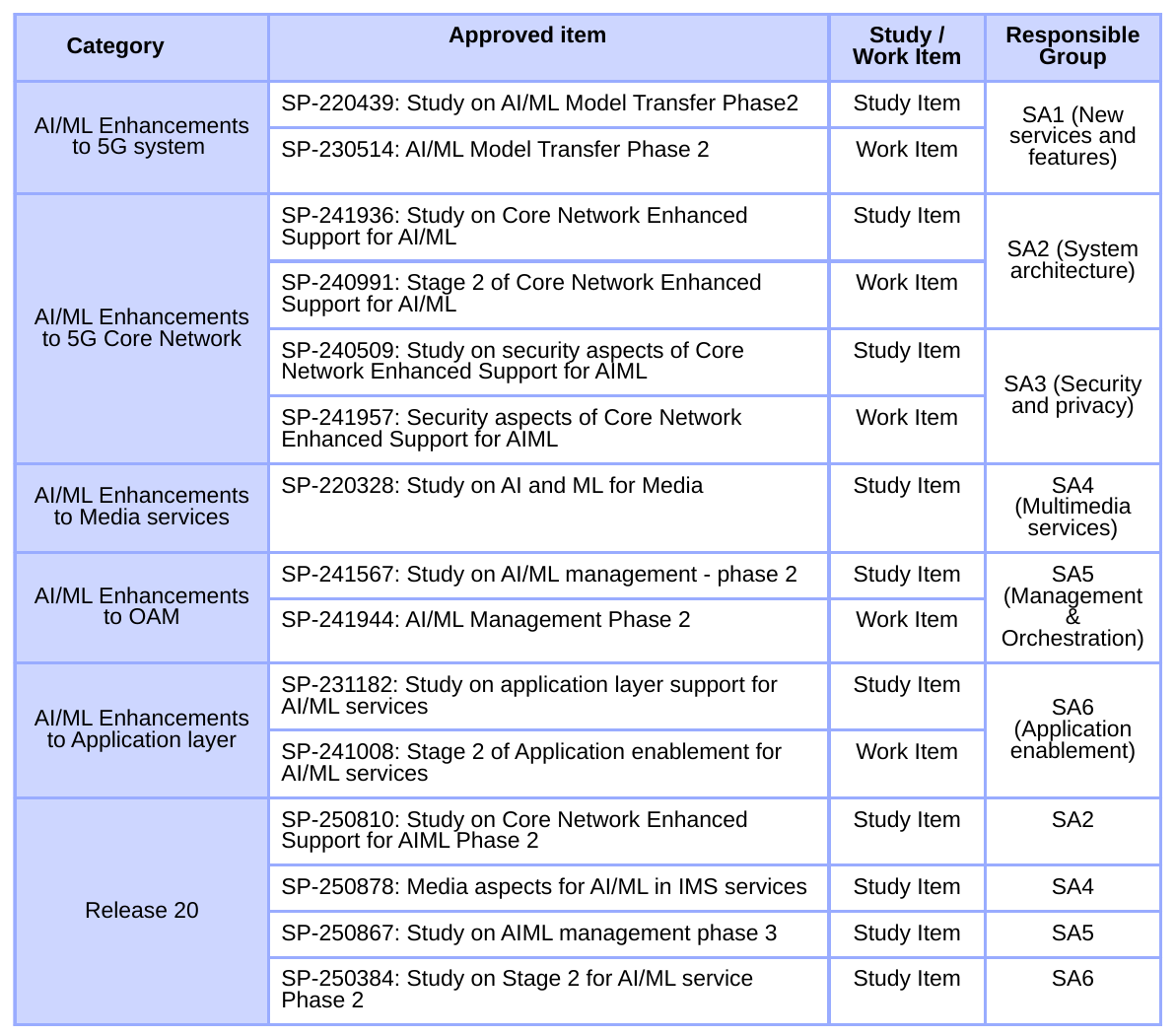}
\caption{Summary of AI/ML approved items in 3GPP Release-19 SA group. While Study Items are feasibility studies for a new feature, Work Items are the concrete specification activities that follow the study phase. The specifications related to these items can be found at https://www.3gpp.org/.} \label{aiml_table}
\end{table*}
The primary focus in Release 19 was on supporting the following features:
\begin{itemize}
    \item Enhancement to location services using AI/ML 
    \item Enable 5G system to support AI/ML operations with the help of NWDAF for VFL
    \item Enable 5GS to use AI/ML to aid policy control along with QoS improvement
    \item Enable 5GS to use AI/ML to foresee abnormal or excessive signalling in the network
\end{itemize} 
NWDAF is the NF in core network which is primarily involved in supporting all the features mentioned above. Its functionalities include data collection from NFs, AFs and OAM, exposure of data analytics, ML model training, support Horizontal FL (HFL) and Vertical FL (VFL). NWDAF is made up of two functions, Model Training Logical Function (MTLF) and Analytics Logical Function (AnLF). MTLF primarily trains the ML models whereas AnLF uses the trained ML models for inference and extracting analytics data. AnLF also facilitates exposure of the analytics data. MTLF and AnLF can be part of the same NWDAF or it can be deployed in different NWDAF instances. Although MTLF was introduced in Release 17, it was enhanced to support all the features listed above in Release 19. 

We address each of the features listed above in detail in the following sections:

\subsection{Location Services Support}
Location Management Function (LMF) in CN is responsible for collecting data from UE/RAN, calculating location information and overall management of location services including provision of services to consumer. UE location information can be requested by NFs in the CN, or Application Functions (AFs). AI/ML based positioning methods have been introduced in RAN to improve accuracy of positioning in Release 18. In Release 19, these methods were integrated in CN architecture at LMF and NWDAF. In these methods, the ML model can be distributed at UE or RAN node or LMF or NWDAF. Fig.~\ref{lmf} provides an overview of UE and RAN node assisted positioning with ML model at LMF. Location services architecture and its functionalities are defined in 3GPP TS 23.273 \cite{ts23273}. The LMF collects input data from UE or RAN to perform location calculation. Using this input data, the ML model can be trained locally at LMF or LMF may request NWDAF to train the ML model. Correspondingly, the performance monitoring for this method can then be performed either at LMF or NWDAF. LMF then calculates location information based on the measurement data by using the ML model. Based on the result, the LMF may change the positioning method or retrain the model.
Privacy and security safe-guards, such as authorization of LMF before it can retrieve an ML model from NWDAF and user consent for collecting data from the UE, must be considered.
The security aspects related to this are covered in 3GPP TS 33.501 \cite{ts33501}.

\subsection{Support for Vertical Federated Learning (VFL)}
Federated Learning uses multiple entities to jointly train an ML model. Vertical Federated Learning (VFL) is a kind of federated learning in which multiple entities train a model collaboratively, without exposing the original data set used by each entity to train the model. Hence, it implicitly maintains the privacy of the raw data. This is an important feature for mobile networks to protect user privacy while supporting AI/ML activities with external entities such as AFs and other operator networks. VFL support in 5G CN enables usage of diverse analytics to train the ML models while maintaining user privacy.

\begin{figure*}[htbp]
    \centering
    \begin{subfigure}[b]{0.45\textwidth}
        \centering    \includegraphics[width=\textwidth]{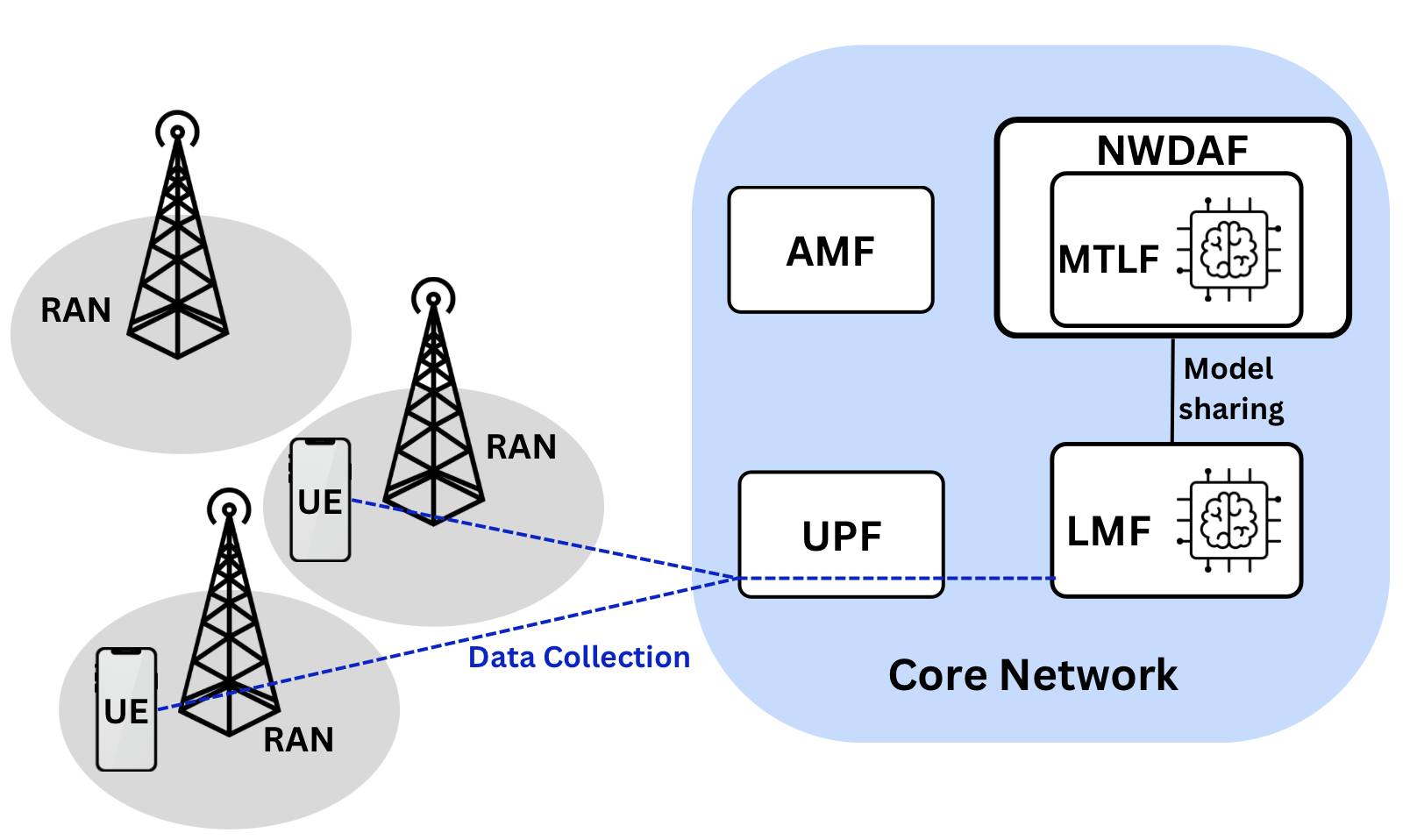}
        \caption{LMF based AI/ML positioning - LMF derives UE location based on inputs from UE and RAN. LMF can get ML models from MTLF of NWDAF}
        \label{lmf}
    \end{subfigure}
    \hfill 
    \begin{subfigure}[b]{0.45\textwidth}
        \centering \includegraphics[width=\textwidth]{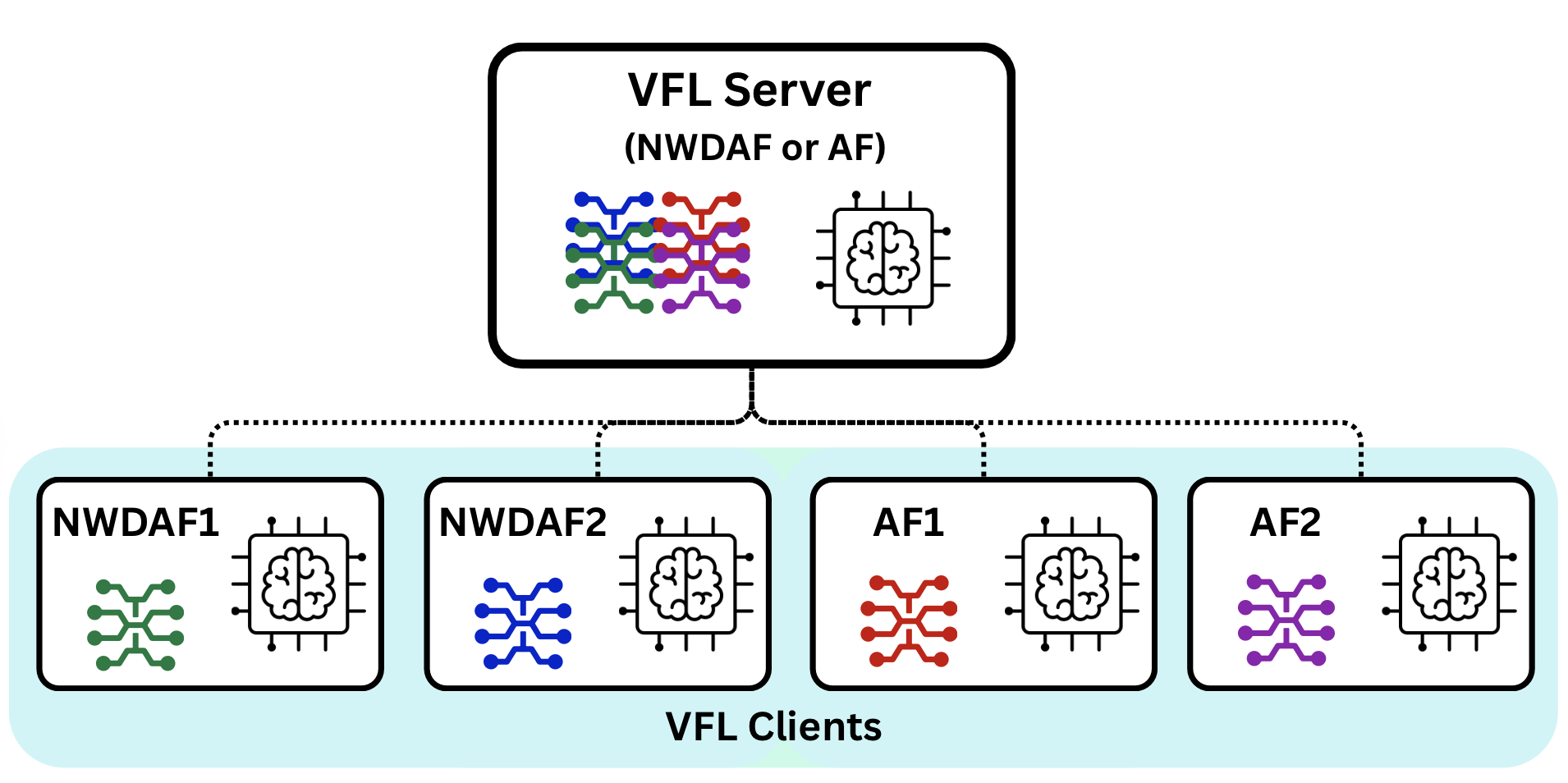}
        \caption{Vertical Federated Learning (VFL) support - NWDAF or AF can be VFL server or VFL clients. VFL server generates the final model based on inputs from VFL clients}
        \label{vfl}
    \end{subfigure}

    \vspace{1em} 

    \begin{subfigure}[b]{0.45\textwidth}
        \centering
        \includegraphics[width=\textwidth]{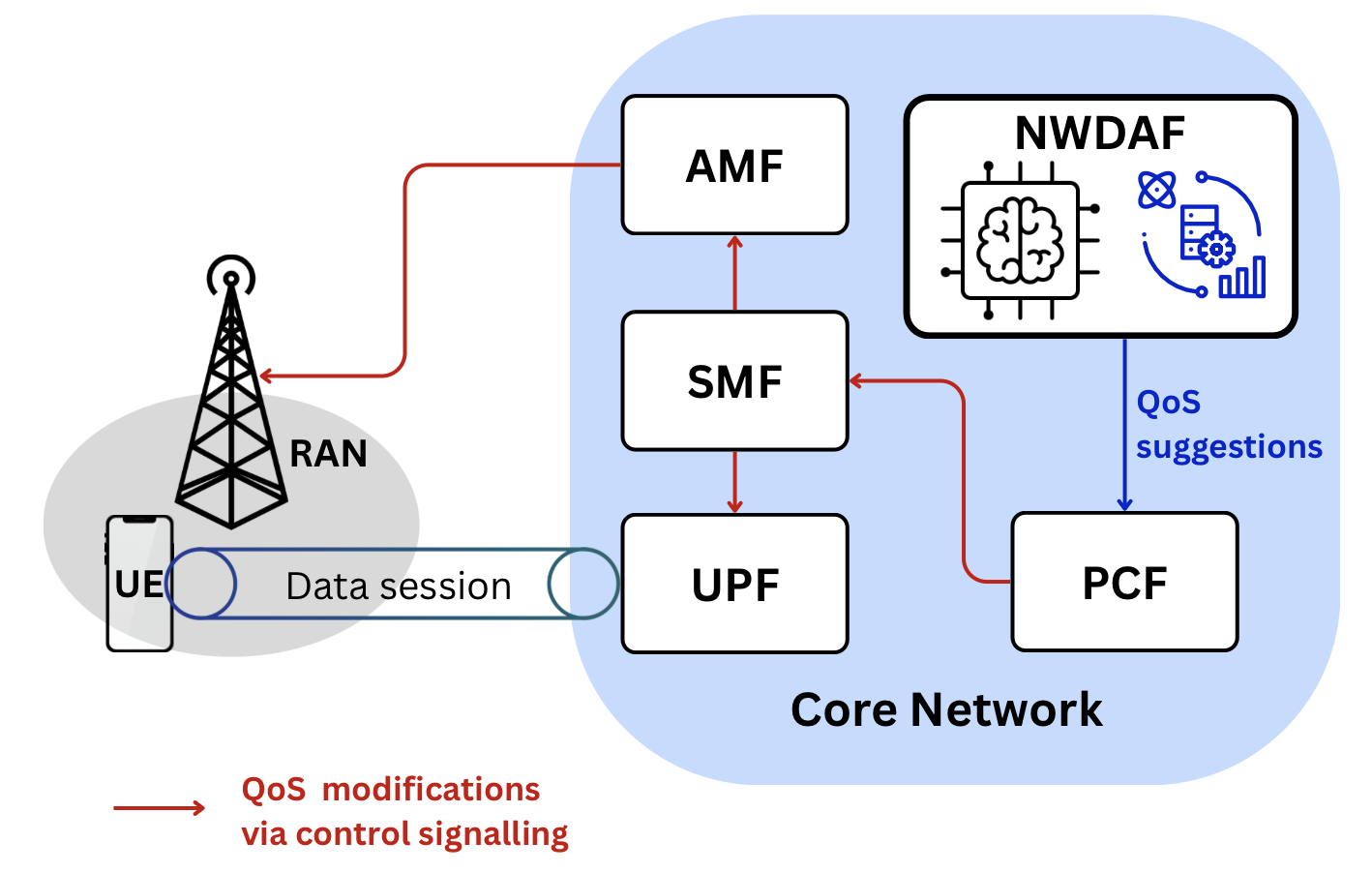}
        \caption{AI/ML based enhancements to QoS policy - NWDAF provides predicted QoS parameters to PCF, it is applied via SMF on UPF and AMF on RAN for the data session}
        \label{qos}
    \end{subfigure}
    \hfill 
    \begin{subfigure}[b]{0.45\textwidth}
        \centering
        \includegraphics[width=\textwidth]{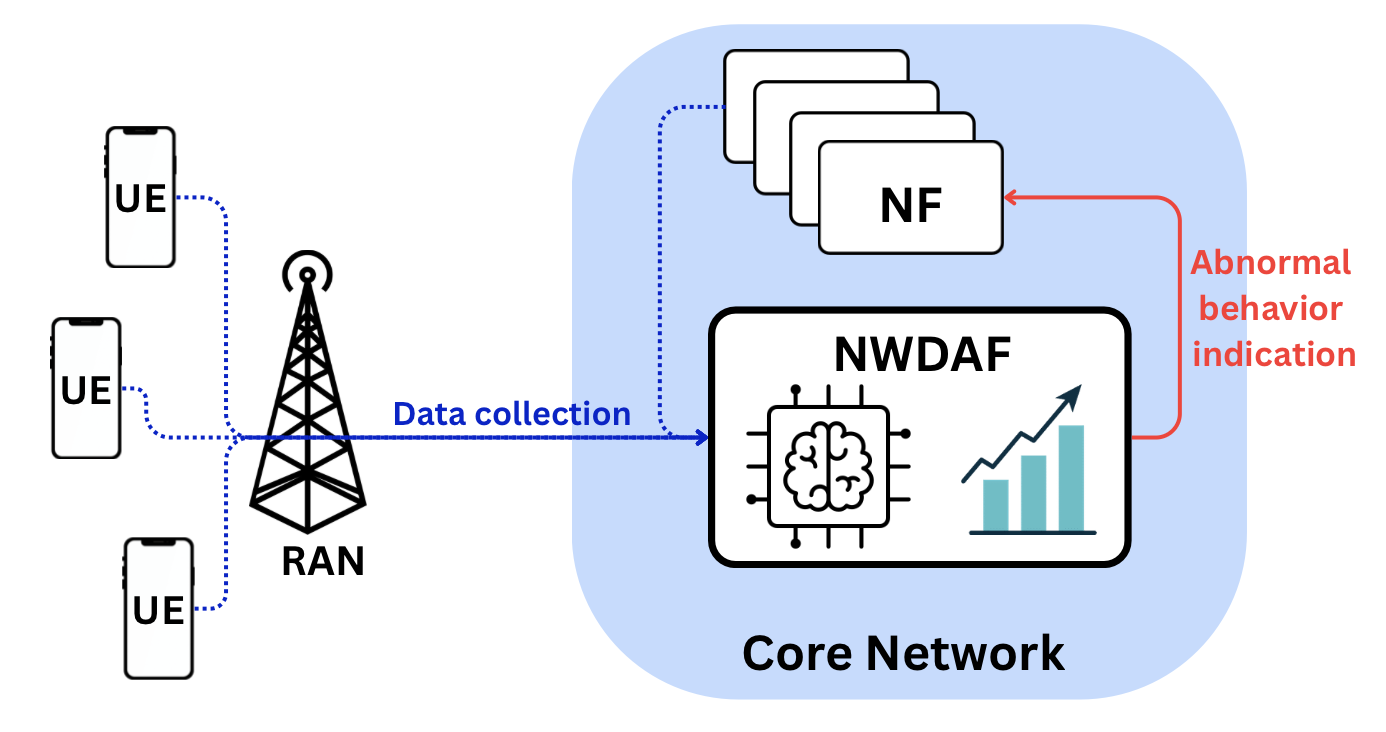}
        \caption{Abnormal behaviour migitation using AI/ML - NWDAF predicts signalling spikes based on inputs from UE, RAN and NFs. Predicitons will be shared with NFs for mitigation }
        \label{abnormal}
    \end{subfigure}

    \caption{AI/ML enhancements to 5G CN.}
    \label{}
\end{figure*}

NWDAF and an AF can train together to develop an ML model (as shown in figure \ref{vfl}). One entity (either NWDAF or AF) can act as a VFL server and other entities (there could be multiple NWDAFs or AFs involved) can act as clients to train the ML model. VFL server requests the VFL clients to train the models locally with the data sets that is available to them and share intermediate results. VFL server then combines the models to derive a final ML model. If the AF is an untrusted AF (AF not hosted by the operator), it can coordinate with NWDAF through Network Exposure Function (NEF) to support VFL interactions. Detailed information flows for VFL training and inference are defined in 3GPP TS 23.288 \cite{ts23288}.

In 3GPP TS 33.501 \cite{ts33501} detailed information flows are described covering authorization of NWDAF or AF to take the role of a VFL server. Information exchange during the VFL process is protected using Transport Layer Security (TLS). 

\subsection{Policy control and QoS enhancements using NWDAF}
Policy Control Function (PCF) in the CN manages policy control and charging framework. Its functionalities include management of UE access and mobility related policies, UE policy information, and data session policies. QoS control is a part of the data session management. Till Release 18, PCF determined the QoS parameters for individual data sessions by considering the service requirements provided by AFs. These values were applied to the data session (service flow) through Session Management Function (SMF). Consequently, PCF would check if the applied parameters satisfied the service requirements (by fetching service experience analytics from NWDAF). If it was not met, it would update the QoS parameters and send it to SMF for reapplying the modified values. PCF would then recheck the new service analytics to determine if the QoS modifications helped. This rechecking and applying would happen several times before the requested service requirements could be met.

In Release 19, the goal was to simplify this process by using AI/ML methods and as shown in Fig.~\ref{qos}, ML based policy control and QoS management support was introduced. NWDAF collects policy related analytics from AFs, 5G NFs and OAM entities to make predictions \cite{ts23288}. PCF can utilize NWDAF’s policy assistance feature providing QoS parameters and requested Quality of Experience (QoE). In response, NWDAF provides possible QoS parameter sets and their corresponding predicted QoE. PCF uses these values to make adjustments and achieve a better QoE.  This ML based QoS and policy assistance can be provided for a single UE or a set of UEs. QoE can be requested for a particular application or applications, which could be associated to single or multiple data session flows in the particular UE or UEs.

\subsection{Abnormal Behaviour Mitigation and Prevention using NWDAF}
Abnormal behaviour in the 5G network here refers to unexpected spikes in signalling load (also referred to as signalling storm). NWDAF was already providing analytics related to NF load information, suspected Distributed Denial of Service (DDoS) attacks, and abnormal behaviour. However, excessive signalling possibly due to Narrowband Internet of Things (NB-IoT) devices were not studied earlier. Combining all these features, the focus in Release 19 was to develop mechanisms to prevent signalling storm with the help of NWDAF analytics. 

As shown in figure \ref{abnormal}, NWDAF \cite{ts23288} learns about abnormal behaviour in the network using a trained ML model. It monitors the network for unusual deviation in signalling levels to identify and learn about the signalling storm. It identifies the sources (NFs and/or UEs) of the excessive signalling. NWDAF monitors successful and failed messaging attempts by UEs, from NFs, load on NFs, and other such relevant data to identify the origin of the signalling storm. Based on its analysis, it can report about impacted NFs, probable reason for the abnormal signalling and its source (UEs or NFs). NWDAF provides this signalling storm analytics to other NFs like Access and Mobility Management Function (AMF), SMF, Unified Data Management (UDM), Network Repository Function (NRF), PCF, OAM and AF. NWDAF can also report about a predicted signalling storm with a validity period. Mitigation of the problem would be based on operator policy or network configuration. 

 Location services support and VFL support, both indicate how AI/ML applications can be supported efficiently when communicating through the network, which is 'Network for AI' usecase. On the other hand, signalling storm mitigation and Policy control handling can be efficiently handled using ML model predictions. These features indicate how AI can be used in the network to improve its management and efficiency, which is 'AI for network'.   

%

\section{AI/ML Enhancements to Media Services}\label{sa4}
%


In Release 19, 3GPP introduces AI/ML models to efficiently support a wide range of media processing applications, spanning from non-real-time use cases like image recognition and speech/face recognition to real-time applications such as live translation, speech-to-text, text-to-speech, object detection, and enhancing quality of video streaming. It has been studied how AI/ML models and data can be distributed across the 5G system or partitioned between different endpoints \cite{tr26927}. To enable AI/ML-based media processing, three model split scenarios have been considered: AI inference at the UE, AI inference at the media server, and split inference between the UE and the media server. AI/ML is also being leveraged to optimize media compression algorithms and to enhance the efficiency of transmission and deployment of AI/ML models and data. This includes the study of compression techniques for both models and data. To support this, new logical functions—AI inference engine, AI data access/delivery, and FL engine—have been introduced at both the UE and the media server. Fig.~\ref{media_example} depicts these new logical functions introduced for enhancing media services. The findings from this study are expected to lead to normative work in Release 20. This feature aligns with the 'Network for AI' paradigm within the 5G system.
\section{AI/ML Enhancements
to Application layer}\label{sa6}
\begin{figure}[htbp]
\centering 
\begin{subfigure} [b]{0.48\textwidth}
    \centering
   \includegraphics[width=\linewidth]{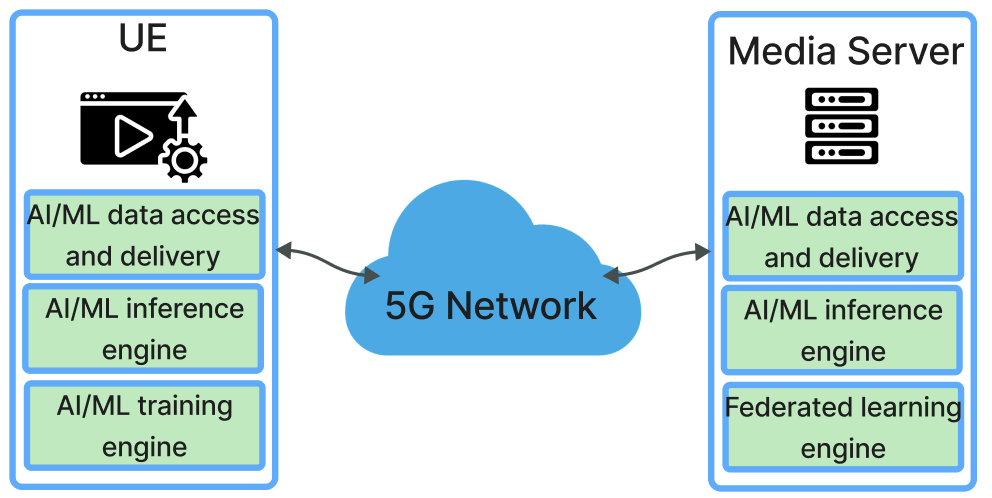}
    \caption{AI/ML enhancements to Media Services}
    \label{media_example}
\end{subfigure}
\hfill 
\vspace{0.05cm}
\begin{subfigure}[b]{0.48\textwidth}
    \centering
    \includegraphics[width=\linewidth]{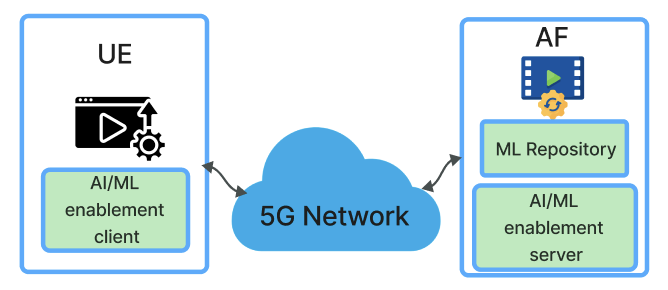}
    \caption{AI/ML enhancements to the Application layer}
    \label{appl_layer}
\end{subfigure}
\caption{AI/ML enhancements to media services and application layer in Release 19. Green boxes indicate AI/ML enhancements.} 
\end{figure}
3GPP introduced the Application layer analytics framework in Release 18 to provide generic exposure and value added services for verticals and Application Service Providers (ASPs) at the application layer. The framework supports analytics related to end-to-end application performance, edge server load, service APIs, location accuracy, and network slice performance and analysis. In Release 19, additional enhancements are proposed to strengthen analytics support through AI/ML techniques, along with extensions to enable UE-to-UE communication at the application layer. An AI/ML enablement layer is introduced at the application layer to facilitate AI/ML operations for one or more vertical applications, either at the UE and/or the AF \cite{ts23482}. The Application layer AI/ML operation
logic is controlled by an AF. The AI/ML enablement layer can be deployed in various configurations: centrally on a cloud platform that aggregates data from multiple edge platforms, locally at the edge platform, or in a hierarchical manner across both edge and central cloud platforms. The hierarchical deployment is particularly beneficial for use cases involving FL, distributed learning, or distributed inference.

Fig.~\ref{appl_layer} depicts the AI/ML enhancements to the application layer. The AI/ML enablement server at the AF manages the AI/ML behavior of vertical applications and assists with tasks related to the delivery and distribution of AI/ML models and data at the AI/ML enablement layer. Additionally, a new ML repository is introduced at this layer to support the storage and retrieval of AI/ML services, including ML model and FL member registration/updates for vertical use cases. 
The introduction of this layer brings support for various AI/ML functions, including lifecycle management, model training and distribution, FL member updates, AI/ML operation management, VFL member selection, transfer learning deployment for HFL, and advanced analytics aimed at improving energy efficiency in the data network. 

This feature is in line with the 'Network for AI' paradigm within the 5G system.

\section{AI/ML Enhancements to Operations and Management}
\label{sa5}

The OAM system is responsible for management, configuration, optimization and maintenance of network resources and services in the mobile network. For effective deployment of AI/ML capabilities in 5G system, it is crucial to manage ML models and AI/ML inference functions according to specific characteristics and requirements of the use cases they support. AI/ML capabilities were introduced in Management Data Analytics (MDA) framework in Release 18. The AI/ML management capabilities include management and operations for ML training, ML Testing, AI/ML emulation, ML entity deployment and AI/ML inference. In Release 19, the MDA framework was extended to support AI/ML management capabilities for advanced learning techniques such as transfer learning, FL, reinforcement learning and Distributed learning \cite{ts28105}. 

The MDA Function (MDAF) offers the MDA management service that allows any authorized consumer to request and receive analytics. 
The main objective of these enhancements was to enable efficient management of AI/ML capabilities within the RAN and 5G Core. Examples of these capabilities include:
\begin{itemize}
    \item \textit{AI/ML based Coverage and Capacity Optimization in RAN} - This feature detects and resolves coverage related issues in RAN. To support this feature, MDAF provides trained ML models to gNodeB (gNB), and gNB uses it for inference. 
    \item \textit{ML model training and AI/ML inference for 5G core} - NWDAF to support ML functionality, it can get trained ML models from MDAF. AI/ML inference feature support can be provided by MDAF based on configuration.   
\end{itemize}
For these features, the AI/ML training and inference functions can be deployed within the RAN/CN or the MDAF or split between the two. These features are primarily aligned towards better optimization of networks using AI/ML (`AI for networks' use case). 

In the next section, we give an overview of the standardization work undertaken beyond Release 19.

\section{AI/ML Beyond Release-19}
Release 19 is nearing completion, while some Release 20 5G Advanced and 6G study items are in their initial stages. Several AI/ML study items are now getting underway in Release 20 within the SA working group, aiming to broaden the scope of AI/ML services to enhance the efficiency of the 5G architecture. A primary area of focus is exploring how 5GC analytics can be leveraged to optimize user-plane performance. Another key focus is sustainability - evaluating AI/ML energy consumption, efficiency, resource utilization across the ML life cycle steps as well as supporting the use of renewable energy sources. Efforts are underway to streamline the process of ML model registration and discovery within the AI/ML management framework.
At the application layer, ML model performance monitoring, correctness evaluation, and assuring service continuity support across multi-operator domains is being studied.

\section{Conclusion}
\label{conc}
We have provided a summary of AI/ML in mobile network systems, along with an overview of AI/ML activities in 3GPP, highlighting key features introduced in Release 19 under different domains. These features have been categorized into two areas: `AI for Networks' and `Network for AI'. 
Incorporating both these paradigms is resource intensive and requires fundamental changes in
network architecture and protocols. However,
considering utility and widespread acceptance of
AI/ML based algorithms, standardization bodies
must provide frameworks that support both the paradigms. 
%
Based on ongoing efforts in 3GPP Release 20 and future 6G systems, we anticipate a strong push for AI/ML integration in future networks.


\section{References}

\bibliographystyle{IEEEtran}
\bibliography{ai_ml.bib}

\section{Acknowledgment}

We thank the Ministry of Electronics and Information Technology (MeitY), Government of India for supporting the project.

\end{document}